\documentclass[a4paper]{interact}
\usepackage{multirow}
\usepackage{epstopdf}% To incorporate .eps illustrations using PDFLaTeX, etc.
\usepackage[caption=false]{subfig}% Support for small, `sub' figures and tables
\usepackage{graphicx}
\usepackage{url}
\usepackage{subcaption}
\usepackage[ruled,vlined]{algorithm2e}
\usepackage[backend=bibtex,style=ieee]{biblatex}
\theoremstyle{plain}% Theorem-like structures provided by amsthm.sty

\theoremstyle{definition}

\theoremstyle{remark}

\begin{document}
%\articletype{ARTICLE TEMPLATE}% Specify the article type or omit as appropriate

%\title{Eyes Reveal Emotion Perception: Insights from Modelling Naturalistic Face Viewing}
%\title{Modelling Gaze during Free-viewing and Grounded Emotion Perception}
%\title{Emotion Recognition and Naturalistic Face-Viewing: An Eye-Tracking Study}

\title{\fontsize{14}{16}\selectfont\textbf{Modeling Face Emotion Perception from Naturalistic Face Viewing: Insights from Fixational Events and Gaze Strategies}}

\author{\textit{Meisam J. Seikavandi\textsuperscript{a}, Maria J. Barrett\textsuperscript{b}, and Paolo Burelli\textsuperscript{a}}\\
\affil{\textsuperscript{a}brAIn lab - IT University of Copenhagen,Denmark} \affil{\textsuperscript{b}IT University of Copenhagen, Denmark}}

% \textbf{Changes and updated}
% \\
% \\
% In this updated paper, several enhancements have been made to our previous work. Firstly, we incorporated microsaccadic events and integrated them with pupil size to enrich the feature set for our tasks. Additionally, we meticulously extracted regions of interest from all presented faces, allowing for a comprehensive investigation into emotional eye-gaze strategies. The updated research also encompasses a holistic statistical analysis across all features, providing valuable insights for emotion analysis. Finally, a novel task has been introduced where machine learning models predict the individual emotion perception performance of participants based on their gaze strategies, including fixational, microsaccadic, and pupillary events. These additions contribute to a deeper understanding of the intricate relationship between eye movements and emotion perception.

\maketitle

\begin{abstract}
Face Emotion Recognition (FER) is essential for social interactions and understanding others' mental states. Utilizing eye tracking to investigate FER has yielded insights into cognitive processes. In this study, we utilized an instructionless paradigm to collect eye movement data from 21 participants, examining two FER processes: free viewing and grounded FER.

We analyzed fixational, pupillary, and microsaccadic events from eye movements, establishing their correlation with emotion perception and performance in the grounded task. By identifying regions of interest on the face, we explored the impact of eye-gaze strategies on face processing, their connection to emotions, and performance in emotion perception. During free viewing, participants displayed specific attention patterns for various emotions. In grounded tasks, where emotions were interpreted based on words, we assessed performance and contextual understanding. Notably, gaze patterns during free viewing predicted success in grounded FER tasks, underscoring the significance of initial gaze behavior.

We also employed features from pre-trained deep-learning models for face recognition to enhance the scalability and comparability of attention analysis during free viewing across different datasets and populations. This method facilitated the prediction and modeling of individual emotion perception performance from minimal observations. Our findings advance the understanding of the link between eye movements and emotion perception, with implications for psychology, human-computer interaction, and affective computing, and pave the way for developing precise emotion recognition systems.
\end{abstract}


\section{Introduction} \label{sec:intro}

Facial emotion recognition (FER) is fundamental to human social interactions, enabling individuals to decipher emotions from facial expressions \cite{walker1998emotions,smith2005transmitting}. While methodologies like brain imaging and physiological signals have been employed to probe this complex process \cite{collin2013facial,valenza2014revealing,bradley2008pupil}, eye tracking emerges as a non-invasive yet insightful tool, offering deep insights into visual attention and emotional processing \cite{skaramagkas2021review}.

Previous research has leveraged eye movements to understand emotion perception (EP) in adults \cite{aracena2015neural, chaby2017gaze}, shedding light on atypical EP associated with conditions such as autism, Attention-Deficit/Hyperactivity Disorder (ADHD), schizophrenia, and certain types of dementia \cite{tsang2018eye,russell2021novel}. Diagnostic eye-tracking tasks, such as the antisaccade task, play a crucial role in identifying diseases like dementia and Alzheimer's \cite{readman2021potential, russell2021eye}.

The shift towards naturalistic tasks, mirroring real-life scenarios, gains attraction due to their relaxed environment, making them suitable for lightweight EP assessments. Eye-tracking research extends its applications to clinical diagnosis and human-robot interaction (HRI), with potential implications for designing socially intelligent robots \cite{fu2022preliminary}.

\begin{figure}[ht]
\vspace{-2em}
\vskip 0.2in
\begin{center}
\centerline{\includegraphics[width=\columnwidth]{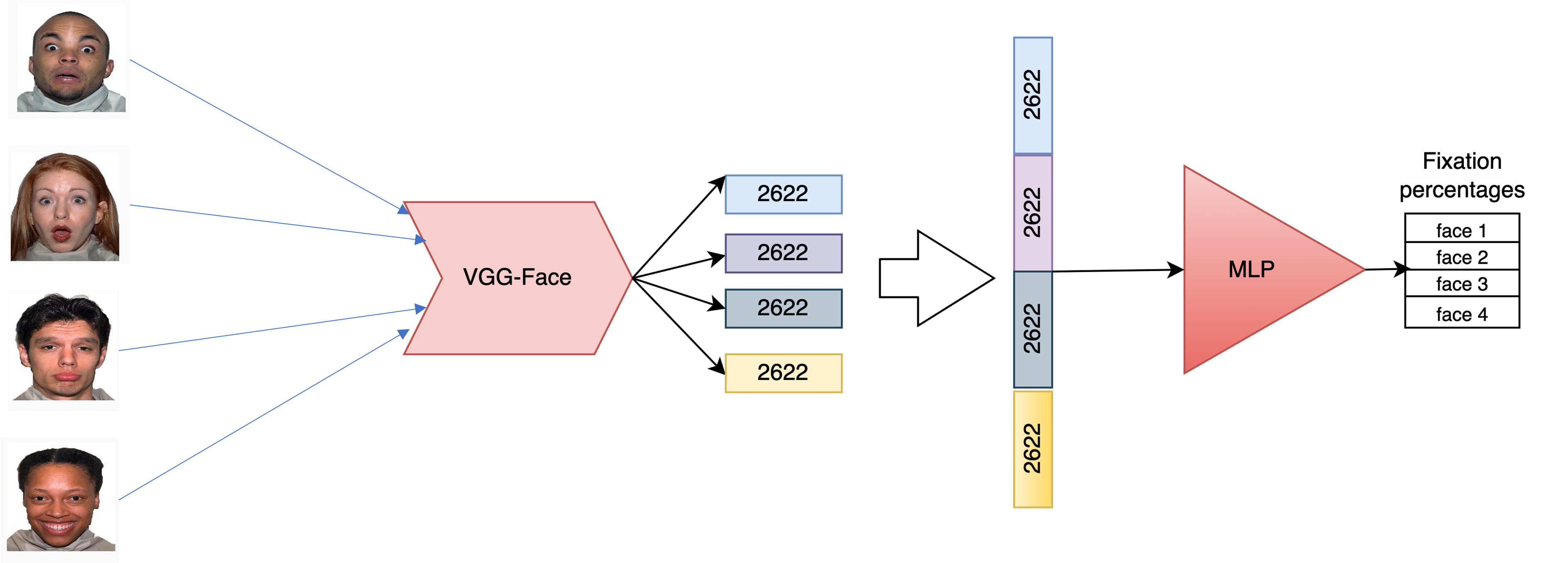}}
\caption{The network structure designed for predicting fixations in Task 2. The model employs a three-layer MLP to analyze spatial, temporal, and spatiotemporal features to predict the dwell time percentage for each face.}
\label{fig:Network}
\end{center}
\vskip -0.2in
\vspace{-1em}
\end{figure}

In this chapter, we extend the work we have done in \cite{seikavandi2023gaze} and delve more into FER using an instructionless paradigm, exploring two FER processes: free viewing and grounded FER. We aim to unravel the intricate relationship between eye movements and emotion perception by scrutinizing fixational, pupillary, and microsaccadic events extracted from eye movement data. By integrating these gaze features with deep-learning models, our study uniquely combines the analysis of microsaccadic events and advanced computational techniques to enhance the understanding of emotional processing.

By delineating regions of interest in the face and harnessing deep-learning models for face recognition, we investigate the role of eye-gaze strategies in face processing and their link to presented emotions and emotion perception performance. Leveraging features extracted from pre-trained deep-learning models, we analyze attention during free viewing, enhancing scalability and comparability across datasets and populations. Furthermore, we employ a sequential model with bidirectional LSTM layers to capture the temporal aspects of fixation between regions of interest, providing insights into the dynamic nature of gaze behavior during face viewing.

Our study contributes to the efficiency of EP assessments by predicting FER success based on gaze features during face viewing. This approach not only enhances the precision and ecological validity of emotion recognition systems but also has significant implications for clinical diagnostics and human-computer interaction. By extending the understanding of the eye movements-emotion perception relationship, our work impacts psychology, HCI, and affective computing domains, offering potential advancements in FER research and applications.

\begin{figure}[ht]
\vspace{-1em}
\vskip 0.2in
\begin{center}
\centerline{\includegraphics[width=0.9\columnwidth]{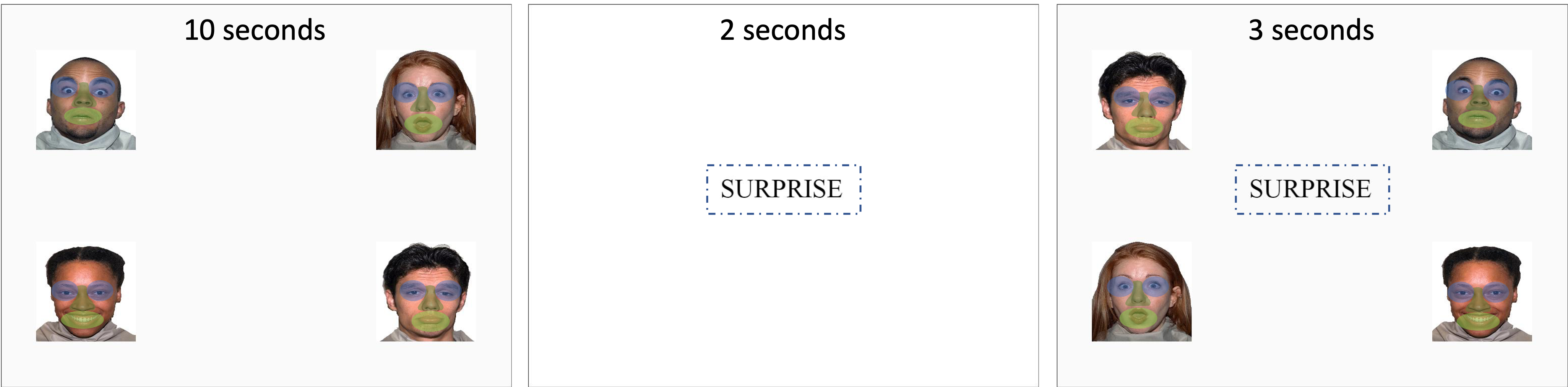}}
\caption{The FER task unfolds in three steps: (1) overlaying areas of interest, (2) presenting faces with emotions, and (3) randomizing locations to study gaze patterns.}
\label{fig:FER}
\end{center}
\end{figure}
\vspace{1em}
\begin{figure}[ht]
%\vspace{-1em}
\vskip 0.2in
\begin{center}
\centerline{\includegraphics[width=0.9\columnwidth]{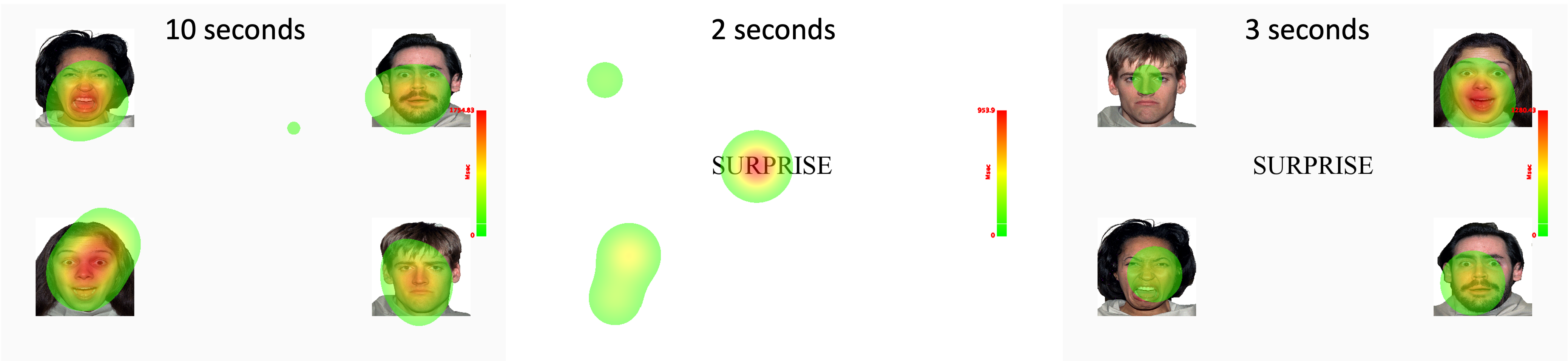}}
\caption{The heatmaps display FER trial with no instructions. Different emotions have distinct fixation distributions, showcasing varied attention patterns across emotions.}
\label{fig:FER_results}
\end{center}
\vskip -0.2in
%\vspace{-1em}
\end{figure}

\section{Background}

\subsection{Instructionless FER Task}
This section outlines the instructionless FER task initially developed by Russell et al. \cite{russell2021novel} and our modifications to enhance the understanding of the FER process.

Russell et al. designed the task to identify early-stage frontotemporal dementia, comprising:
\begin{enumerate}
\item Displaying four faces with distinct emotions for 10 seconds
\item Presenting an emotion word for 2 seconds
\item Showing the word and faces together for 3 seconds
\end{enumerate}
This setup aimed to intuitively direct participants' gaze to the face matching the emotion if they remembered its position. The constant positions of the faces simulated a memory retrieval task, which helped analyze free-viewing and retrieval phases and considered working memory as a potential confounder.

\subsection{Modifications to the Instructionless FER Task}

To deepen our insight into the FER process and mitigate the effect of working memory, we adjusted Russell et al.'s design by randomizing face positions in Step 3. This required participants to recognize rather than recall face positions, effectively isolating the role of memory. This alteration helped distinguish between the free-viewing (Step 1) and grounded FER (Step 3) phases.

Through these changes, we analyzed gaze behavior and performance variations between these phases, generating quantitative data on cognitive processes in FER. These adjustments better simulate real-world scenarios, enhancing our study of the links among eye movements, emotion perception, and attention in FER tasks.

\subsection{Microsaccades}

Microsaccades are small, involuntary eye movements that occur during visual fixation, typically lasting 6-30 milliseconds and with amplitudes under 0.1 degree of visual angle. Interspersed with slow drifts, they prevent retinal image fading and are vital for tasks requiring sustained visual attention, like reading \cite{howard2017temporal}. Studies suggest that microsaccadic activity responds to emotional stimuli, affecting attention and emotion-related cognitive processes. Emotional arousal, fatigue, and saccade preparation can influence microsaccades, altering their rate and magnitude depending on the emotional context \cite{kashihara2020microsaccadic}. While some research reports significant changes in microsaccadic behavior in response to different emotions, others find no noticeable differences \cite{strauch2018pupil}. Further studies link microsaccades with cognitive effort, affective priming, and arousal, highlighting their sensitivity to these factors and their role in emotional and cognitive processing \cite{strauch2018pupil}. This underlines the importance of microsaccades in understanding visual attention and emotion perception.

\begin{figure}
\centering
\subfloat[step 1]{%
\resizebox*{5cm}{!}{\includegraphics{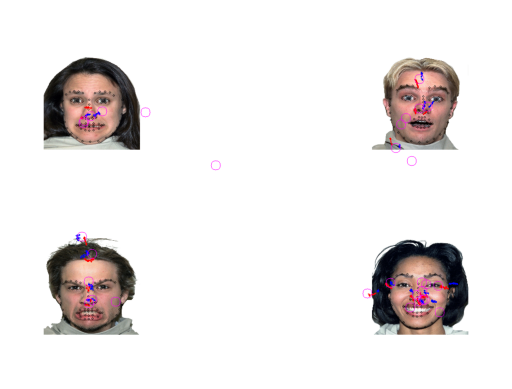}}}
\subfloat[step 2.]{%
\resizebox*{5cm}{!}{\includegraphics{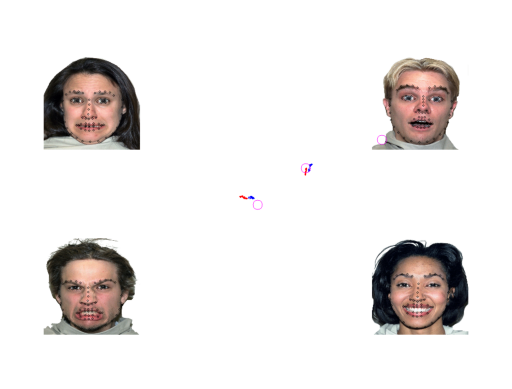}}}
\subfloat[step3]{%
\resizebox*{5cm}{!}{\includegraphics{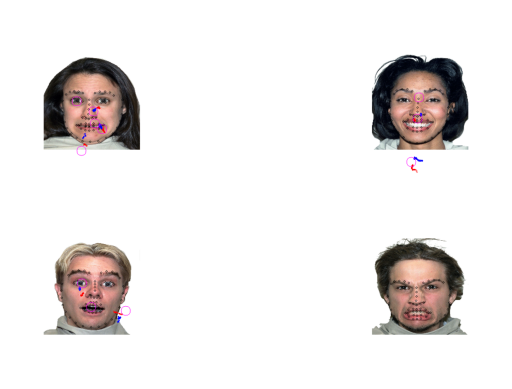}}}\hspace{5pt}
\caption{Microsaccades across different steps. Step 1 shows baseline microsaccadic activity, Step 2 indicates changes in response to emotion words, and Step 3 highlights microsaccades during emotion recognition.} 
\label{fig:sample-figure}
\end{figure}

\subsection{Eye Gaze Strategies}
% Research on eye gaze strategies for emotion perception provides intriguing insights, revealing variations based on factors such as age \cite{chaby2017gaze}, gender \cite{coutrot2016face}, and more. 
Studies indicate that eye movements during emotion recognition in faces follow both stimulus-driven and goal-driven perceptual strategies \cite{rodger2023developmental}. Different facial regions contain varying levels of useful information for distinguishing emotions. For example, joyful faces may draw attention to the lips, while sad faces may attract attention to the eyes. These fixation patterns are influenced by attention to the most diagnostic regions of the face for each emotion, indicating a goal-driven influence on gaze patterns \cite{schurgin2014eye}. Furthermore, gaze direction can modulate emotion perception in facial expressions, influencing how emotions are perceived based on gaze direction. A study found that faces with averted gaze were rated higher overall in terms of perceived likelihood of experiencing emotion compared to direct gaze faces, demonstrating an interaction between gaze direction and perceived emotional disposition \cite{liang2021emotional}.

\section{Data Collection}
\label{sec:data collection}

\subsection{Participants} We recruited 21 volunteers with normal or corrected vision using glasses or lenses who reported no history of attention deficits or cognitive impairments. One participant was excluded from the analysis due to missing information in some trials. The remaining 20 participants had completed education ranging from high school to PhD, with MSc being the mode. Table \ref{tab:particpants} provides an overview of other demographics.
The information statement form, which was approved by the legal department of our institution, was signed by all participants. 
\begin{table}[t]
    %\centering
    \caption{Participant characteristics }
    \begin{center}
        %\begin{small}
        \begin{sc}
        \begin{tabular}{lcc}
        \toprule
        & \textsc{Age}  & \textsc{Drift Error}\\%  &  &Education  \\
        \midrule
        %\midrule
         \textsc{Count} & 20 &  20\\% &\textsc{PhD} &2 \\ 
         \textsc{Range} & 23--44 &  0.01--1.21 \\%& \textsc{MSc} & 14\\
         \textsc{Mean} & 29.3 &  0.41\\ %& \textsc{BSc} & 2\\
         \textsc{SD} & 5.3 &  0.28\\% &  \textsc{Diploma}& 2\\
        \bottomrule
        \end{tabular}
        \end{sc}
        %\end{small}
        \end{center}
        %\vskip -0.1in
\label{tab:particpants}
\end{table}

\subsection{Apparatus}
We utilized the Eyelink 1000 Plus eye tracker from SR Research for recording eye movements during FER tasks. Participants were seated in a dark room with their chin stabilized on a chin rest. An 18-inch high-resolution display screen with a resolution of 1024 x 768 pixels was placed 70 cm from the participants. Before each experiment, we performed a 9-point calibration for the eye tracker, followed by a drift correction between trial rounds to ensure accuracy. Recalibration was done whenever the accuracy dropped below the desired threshold.

\subsection{Stimuli}
The NimStim face emotion dataset \cite{tottenham2009nimstim} was employed, generating 60 trials involving four emotive facial images and one emotion word per trial, aiming for a balanced representation of target emotions and diversity in facial identities. We randomized the positions of the faces and the target face to promote effective FER. Each face, sized 200 x 200 pixels, was placed towards the screen corners. The emotion word appeared in a 40-point Times New Roman font, centrally positioned. A fixation cross was shown for 200 ms at the start of each trial to center participant attention.

\subsection{Areas of Interest}
Areas of interest included the four facial images and their corresponding words per trial. Within each face, we defined subareas for the eyes, nose, and mouth based on a CNN-based landmark detection model \cite{zadeh2017convolutional}. These landmarks helped segment the faces into seven regions: mouth, both eyebrows, both eyes, nose, and jaw, further grouped into three categories: eye, nose, and mouth regions.

\subsection{Experiment Protocol}
Participants naturally viewed the screen without specific instructions. After six preliminary trials for acclimation, they completed two rounds of 27 trials each, with a short break in between. The total duration was about 15 minutes. Post-experiment, participants could opt to complete the Reading Minds in the Eyes test online to assess their emotion decoding ability. Fifteen participants took this test, with their performance detailed in Table \ref{tab:particpants}.

\subsection{Preprocessing and Cleaning}
Data from both eyes were collected, but fixation detection relied on the eye with the highest accuracy, confirmed through drift checks. For other analyses, we used binocular data or eye averages. The first six trials were excluded to eliminate initial bias. Fixations were assigned to the nearest area of interest, ensuring clear and consistent data interpretation across trials, simplifying analysis and enhancing data reliability.

\subsection{Microsaccade Extraction Algorithm}

To extract microsaccades from eye-tracking data, we implemented an algorithm that processes velocity and acceleration thresholds to identify significant eye movements. The algorithm preprocesses the data for both left and right eyes, extracts relevant features, and iterates through the samples to detect microsaccades based on predefined velocity and acceleration thresholds. Detected microsaccades are validated by their duration and spatial displacement criteria. The complete algorithm is outlined in Algorithm \ref{alg:microsaccade}.

\begin{algorithm}[H]
\caption{Microsaccade Extraction}
\label{alg:microsaccade}
\SetAlgoLined
\KwIn{Eye-tracking data $df$, fixation index $value$}
\KwOut{Microsaccade count $ms\_count$, left eye gaze lists $gaze\_l\_list\_all$, right eye gaze lists $gaze\_r\_list\_all$}
\SetKwInOut{Parameters}{Parameters}
\Parameters{$velocity\_threshold = 15$, $acc\_threshold = 5000$, $min\_duration = 10$, $max\_duration = 100$}

Preprocess data for left and right eyes: $cleaned\_data\_l$, $cleaned\_data\_r$\;
Extract velocity, acceleration, and gaze data for both eyes\;
Initialize variables: $in\_ms \gets False$, $ms\_count \gets 0$, $ms\_duration \gets 1$\;
Initialize sums: $v\_l\_sum \gets [0.0, 0.0]$, $v\_r\_sum \gets [0.0, 0.0]$, $a\_l\_sum \gets [0.0, 0.0]$, $a\_r\_sum \gets [0.0, 0.0]$, gaze lists\;
Set sample counters: $count \gets 100$, $samples\_len$, $samples\_count \gets 0$\;

\ForEach{sample of $vel\_l, acc\_l, vel\_r, acc\_r, gaze\_r, gaze\_l$}{
    $samples\_count \gets samples\_count + 1$\;
    \If{$samples\_count + 100 > samples\_len$}{
        \textbf{break}\;
    }
    \While{$count > 1$}{
        $count \gets count - 1$\;
        \textbf{continue}\;
    }
    Accumulate $v\_l, v\_r, a\_l, a\_r$ into sums\;
    Append $gaze\_r, gaze\_l$ to respective lists\;
    \If{mean velocities and accelerations exceed thresholds}{
        \If{$in\_ms$}{
            $ms\_duration \gets ms\_duration + 1$\;
            \textbf{continue}\;
        }\Else{
            $in\_ms \gets True$\;
        }
    }\Else{
        \If{$min\_duration \leq ms\_duration \leq max\_duration$}{
            \If{distance between start and end gaze positions is within range}{
                $ms\_count \gets ms\_count + 1$\;
                Reset $count$\;
                Append gaze lists to $gaze\_list\_all$\;
            }
        }
        Reset variables and lists\;
    }
}
\KwRet{$ms\_count, gaze\_l\_list\_all, gaze\_r\_list\_all$}\;
\end{algorithm}

\begin{table}[t]
        \caption{Dwell time \% per step wrt. main areas of interest (target face, non-target face, and word) and D(well) T(ime) C(hange) across emotions.}
        %\vskip 0.15in
        %\vspace{-2em}
        %\hspace{-3cm}
        \begin{center}
        %\begin{small}
        \begin{sc}
     %\resizebox{.45\textwidth}{!}{
        \begin{tabular}{lc rrr rrr r}
        \toprule
        & \textsc{Step 1} & \multicolumn{3}{c}{\textsc{Step 2}} & \multicolumn{3}{c}{\textsc{Step 3}} & DTC\\
        \midrule
        && \multicolumn{2}{c}
        {\textsc{target}} & \textsc{word} & \multicolumn{2}{c}{\textsc{target}} & \textsc{word} \\
        && \textsc{no} & \textsc{yes} && \textsc{no} & \textsc{yes} & &\\
        \midrule
         \textsc{angry} & 23.8 & 7.3 & 10.9 & 68.6 &15.8 & 42.2 & 11.9 & 27.8\\ 
         \textsc{disgust} & 23.5 & 7.7 & 11.1 & 69.8 & 14.9 & 45.3 & 11.4 & 31.7\\
         \textsc{fear} & 25.3 & 7.5 & 8.6 & 69.0 & 17.8 & 34.9 & 10.8 & 16.0\\
         \textsc{happy} & 21.4 & 6.7 & 11.0 & 68.2 & 14.4 & 43.8 & 12.1 & 34.2\\
         \textsc{sad} & 23.0 & 6.2 & 11.4 & 65.8 & 14.8 & 38.5 & 11.3 & 25.1\\
         \textsc{surprise} & 23.9 & 7.3 & 8.4 & 68.9 & 16.8 & 40.9 & 11.5 & 26.1\\
         \midrule
         \textsc{avg.} & 23.5 & 7.1 & 10.2 & 68.4 & 15.7 & 40.9 & 11.5 & 26.8\\
        \bottomrule
        \end{tabular}
    %}
    \end{sc}
    %\end{small}
    \end{center}
    %\vskip -0.1in
    %\vspace{-1em}
    \label{tab:gaze_dist}
\end{table}

\begin{figure}
\centering
\subfloat[fixation distribution on step 1]{%
\resizebox*{5cm}{!}{\includegraphics{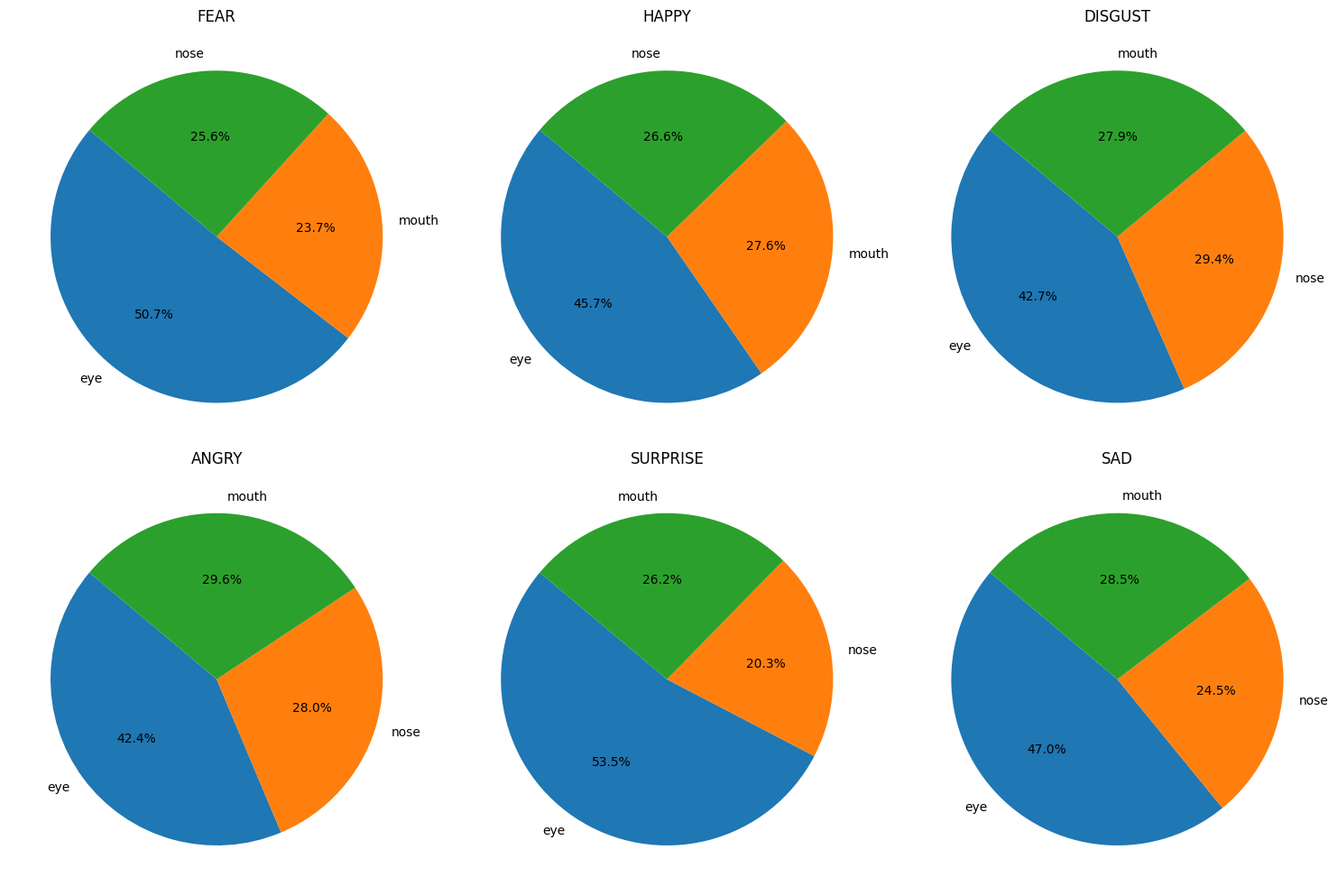}}}
\subfloat[fixation distribution on step 3.]{%
\resizebox*{5cm}{!}{\includegraphics{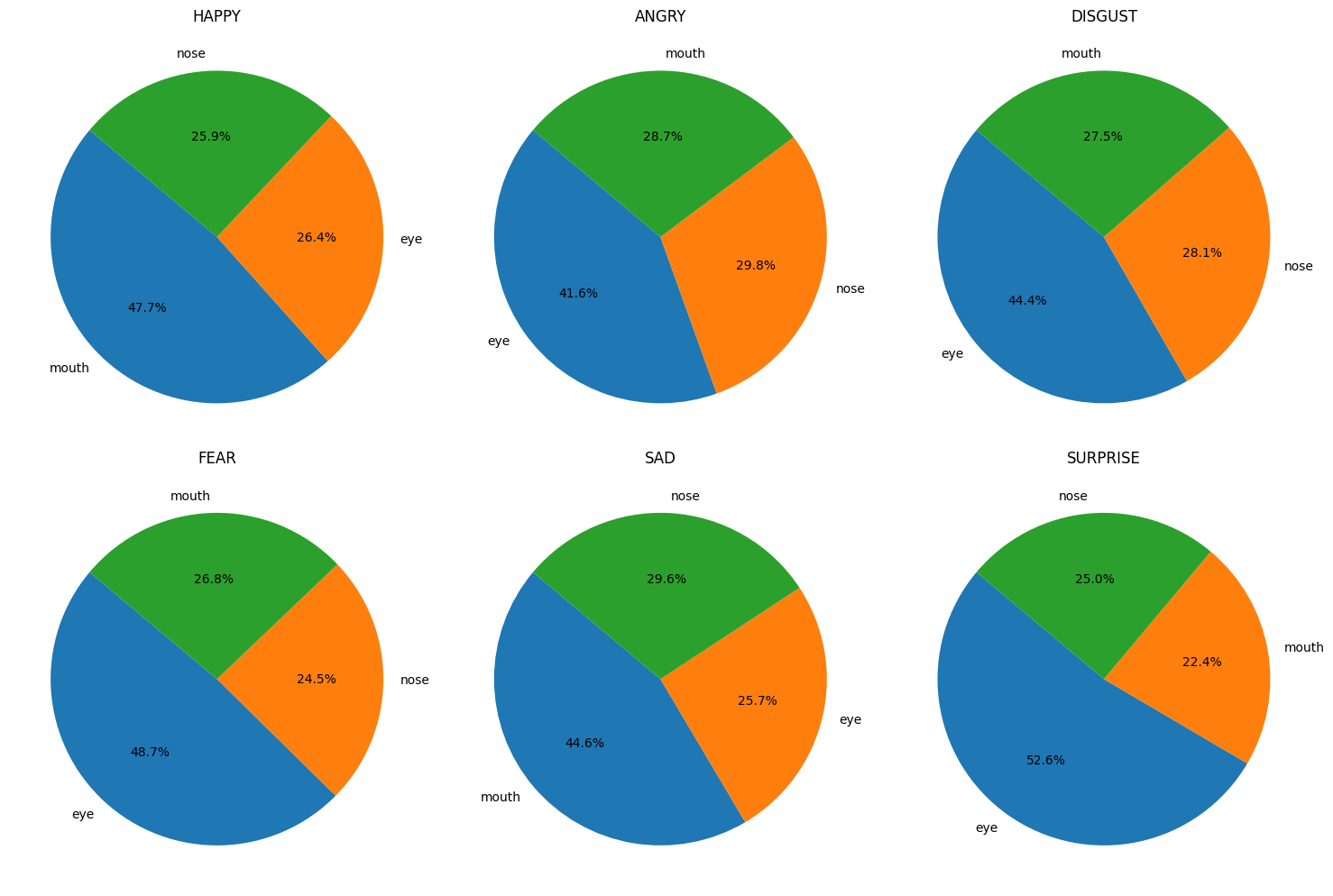}}}\hspace{5pt}
\subfloat[fixation distribution on step 3 where participant looked at target faces]{%
\resizebox*{5cm}{!}{\includegraphics{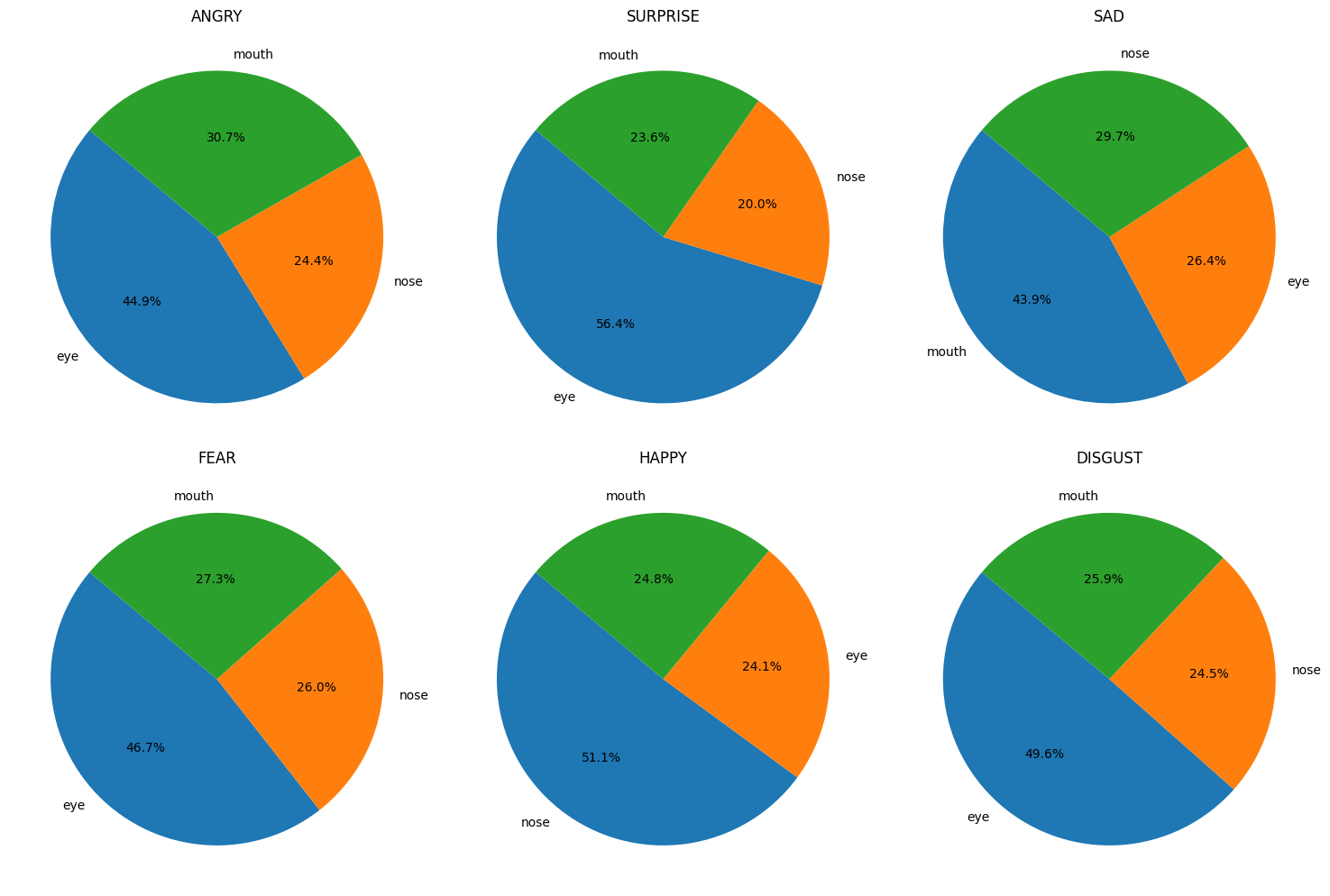}}}\hspace{5pt}
\subfloat[fixation distribution on step 3 where participant looked at non-target faces]{%
\resizebox*{5cm}{!}{\includegraphics{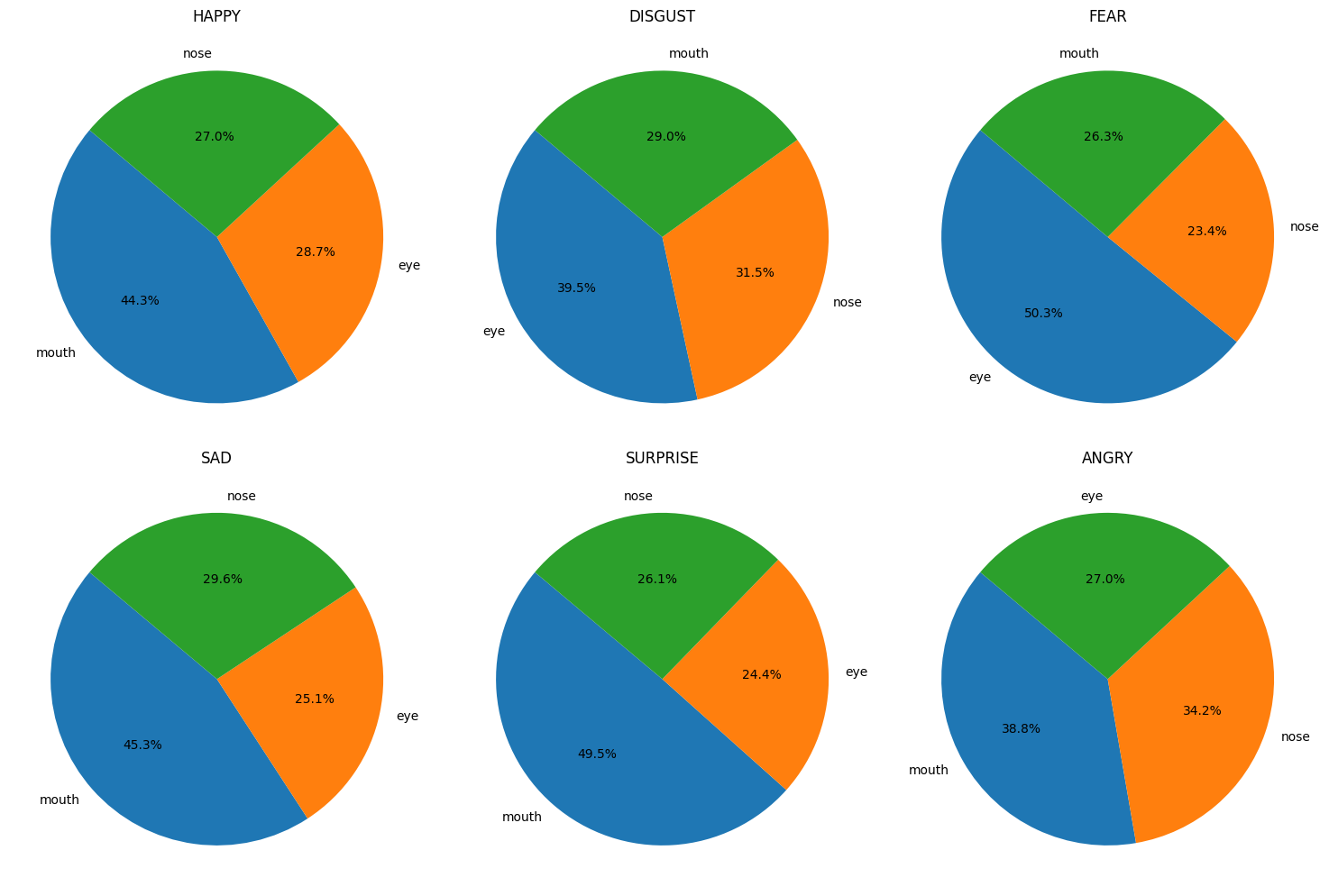}}}
\caption{Fixation distributions over different face regions (eye, nose, and mouth) across the different steps. These distributions help in understanding the attention strategies employed by participants during the FER tasks.} 
\label{fig:eye_mouth_nose}
\end{figure}

\section{Statistical Analysis}

\subsection{Overview}
Following the guidelines suggested by Skaramagkas et al. \cite{skaramagkas2021review}, we employed the dwell time percentage (dwell time \%) as our primary measure for assessing visual attention. This metric calculates the focus duration on a specific area of interest (AOI) as a percentage of the total time spent on a particular step. The change in dwell time for target faces, as proposed by Russell et al. \cite{russell2021novel}, was used to gauge emotion perception (EP) performance:
% \begin{equation}
%     \text{dwell time change} = \text{dwell time \% at step 3} - \text{dwell time \% at step 1}
% \end{equation}

\begin{small}
\begin{displaymath}
dwell\ time\ change = dwell\ time\ \%\ step\ 3 - dwell\ time\ \%\ step\ 1
\end{displaymath}
\end{small}

Results indicate a pattern where participants significantly focus more on the target face after the emotion word is presented, aligning with findings from previous studies \cite{tottenham2009nimstim, russell2021novel, polet2022eye}.

\subsection{Analysis Methods}
Chi-Square and ANOVA tests were used to analyze the relationship between categorical and numerical variables respectively, with Bonferroni correction applied to control for the risk of Type I errors across multiple comparisons \cite{montgomery2021introduction, bonferroni1936teoria}.

\subsection{Significant Findings}
\begin{itemize}
    \item \textbf{Target Emotion:} Variations in fixation duration and microsaccade activity highlighted differences influenced by the emotion depicted on the target faces.
    \item \textbf{Face Regions:} Significant disparities in microsaccade rates across different facial regions underscore their importance in analyzing facial emotions.
    \item \textbf{Interest Periods:} Observable differences in microsaccades and pupil size among different steps suggest varying cognitive demands, which can inform more nuanced analyses in emotion perception studies.
    \item \textbf{Participants:} The diversity in participant responses to identical stimuli illustrates the challenge in crafting a universally applicable model for emotion perception but also highlights the potential of personalized data analysis.
\end{itemize}

\subsection{Performance Analysis}
The analysis revealed that fear generally resulted in lower performance scores, whereas happiness was associated with higher scores. This variability demonstrates differences in the efficiency of emotion recognition among participants.

\subsection{Experimental Observations}
% \begin{itemize}
%     \item \textbf{Step 2:} The focus on the emotion word and the position of the target face suggests a reliance on memory to match the emotion.
%     \item \textbf{Step 3:} There was a significant difference in attention levels between target and non-target faces, with non-target faces displaying fear and surprise attracting considerable focus.
% \end{itemize}

\begin{figure}[ht]
\vspace{-1em}
\vskip 0.2in
\begin{center}
\centerline{\includegraphics[width=0.9\columnwidth]{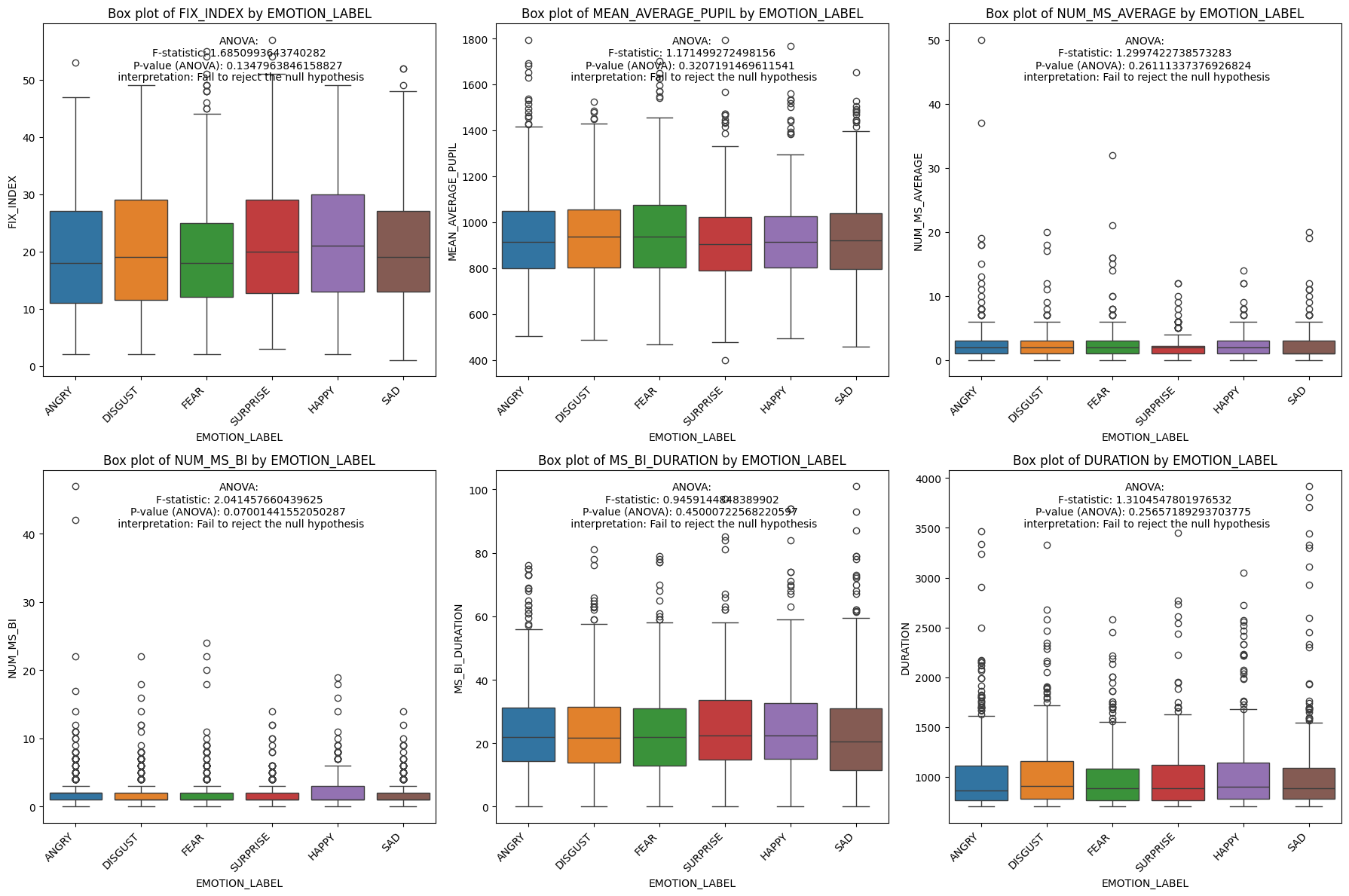}}
\caption{Distribution of fixational, microsaccadic, and pupillary events across different emotions, showing the variability and potential connections between these events and emotion perception.}
\label{fig:emotion}
\end{center}
\end{figure}

\begin{figure}[ht]

\begin{center}
\centerline{\includegraphics[width=0.9\columnwidth]{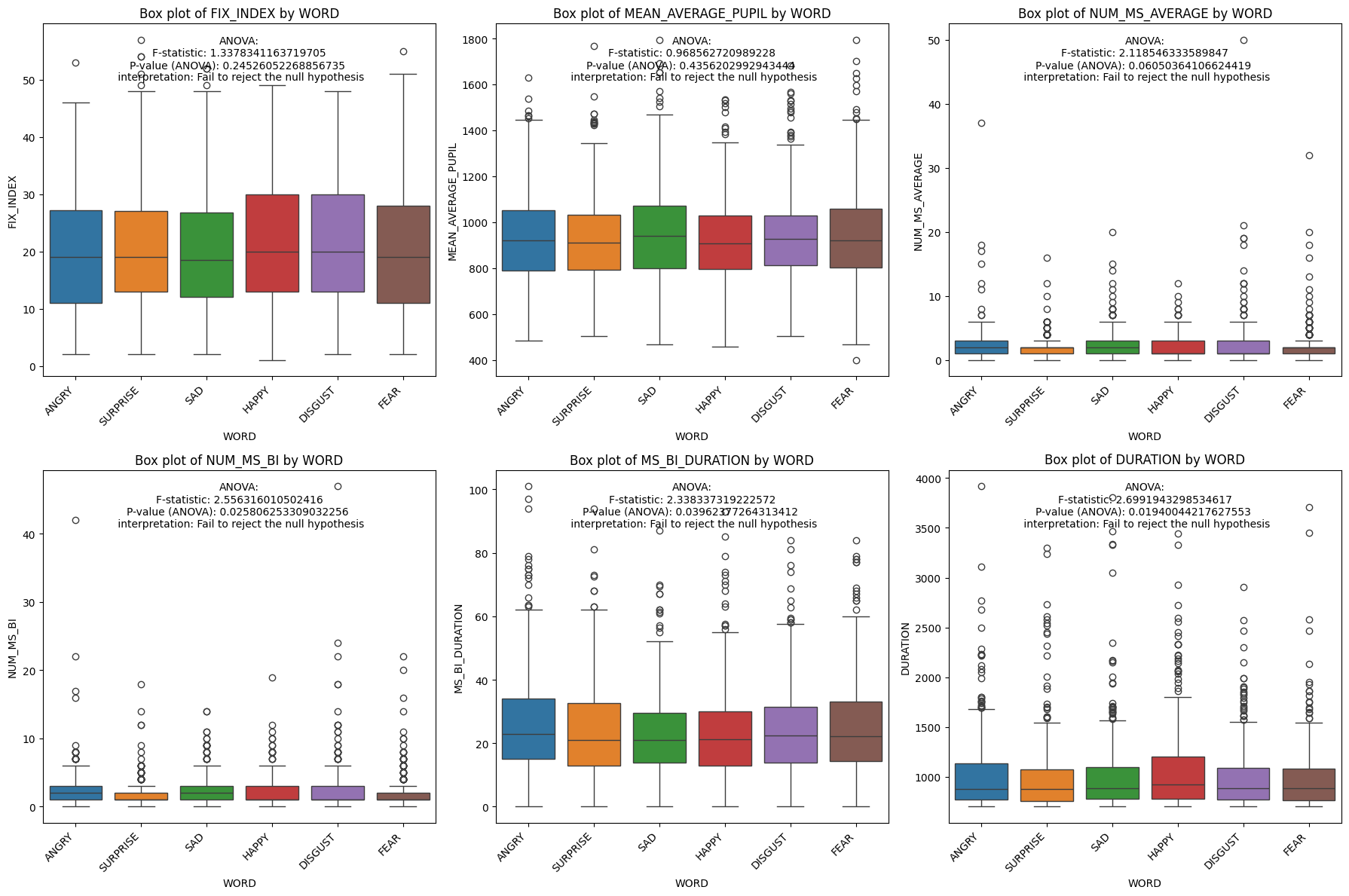}}
\caption{Comparison of fixation duration, microsaccade rate, and duration across different target emotions. The results suggest emotional differences influence these eye-tracking metrics.}
\label{fig:Target}
\end{center}
\end{figure}

% \begin{figure}[ht]
% \begin{center}
% \centerline{\includegraphics[width=0.9\columnwidth]{boxplots/Landmark labels.png}}
% \caption{Face landmarks }
% \label{fig:box_land}
% \end{center}
% \end{figure}

% \begin{figure}[ht]
% \begin{center}
% \centerline{\includegraphics[width=0.9\columnwidth]{face_region_1.png}}
% \caption{Face region 1 }
% \label{fig:FER}
% \end{center}
% \end{figure}

\begin{figure}[ht]
\begin{center}
\centerline{\includegraphics[width=0.9\columnwidth]{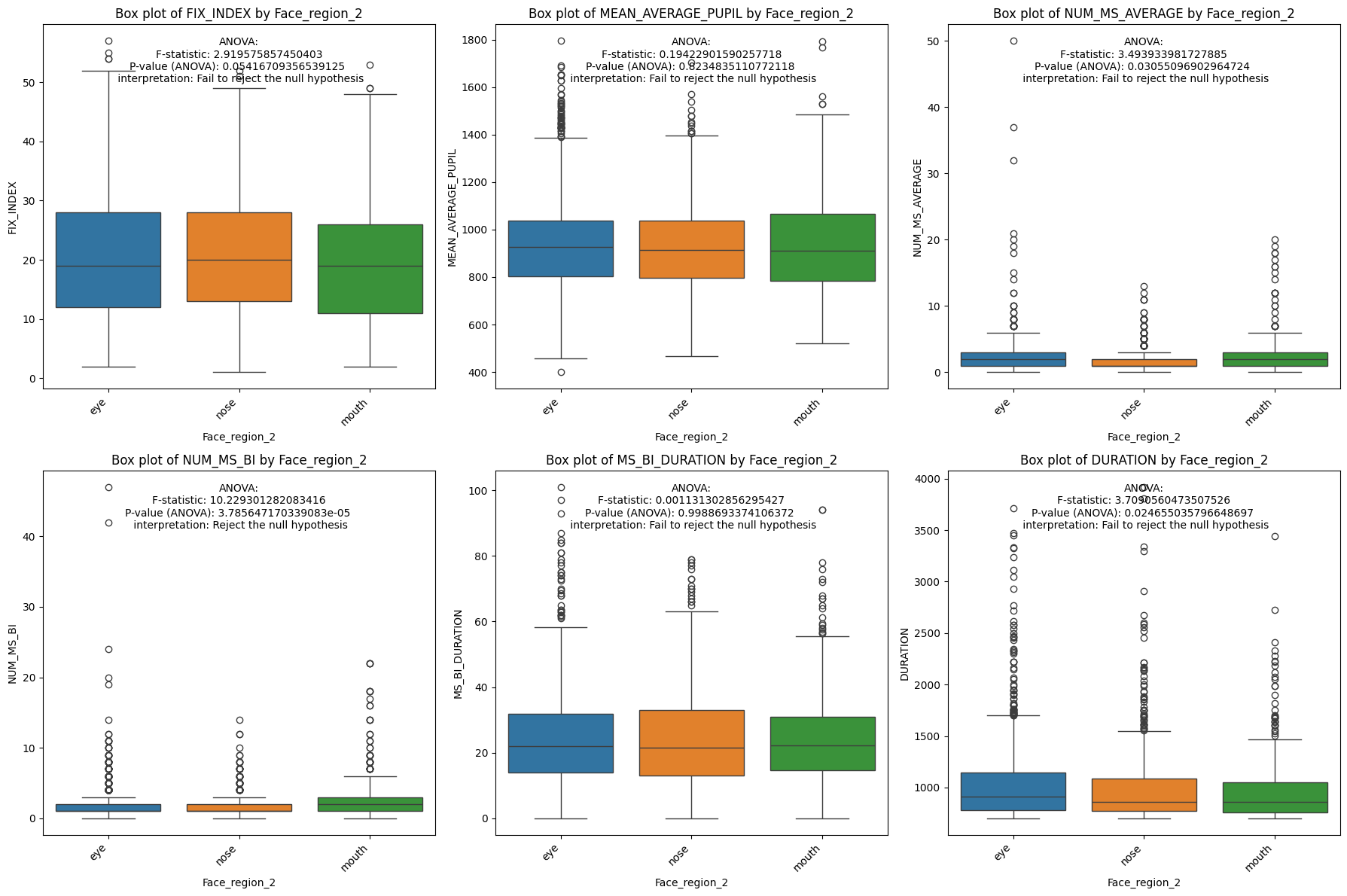}}
\caption{Microsaccade rate variations across different face regions, highlighting the significance of specific facial areas in emotion perception tasks.}
\label{fig:face_regions}
\end{center}
\end{figure}

\begin{figure}[ht]
\begin{center}
\centerline{\includegraphics[width=0.9\columnwidth]{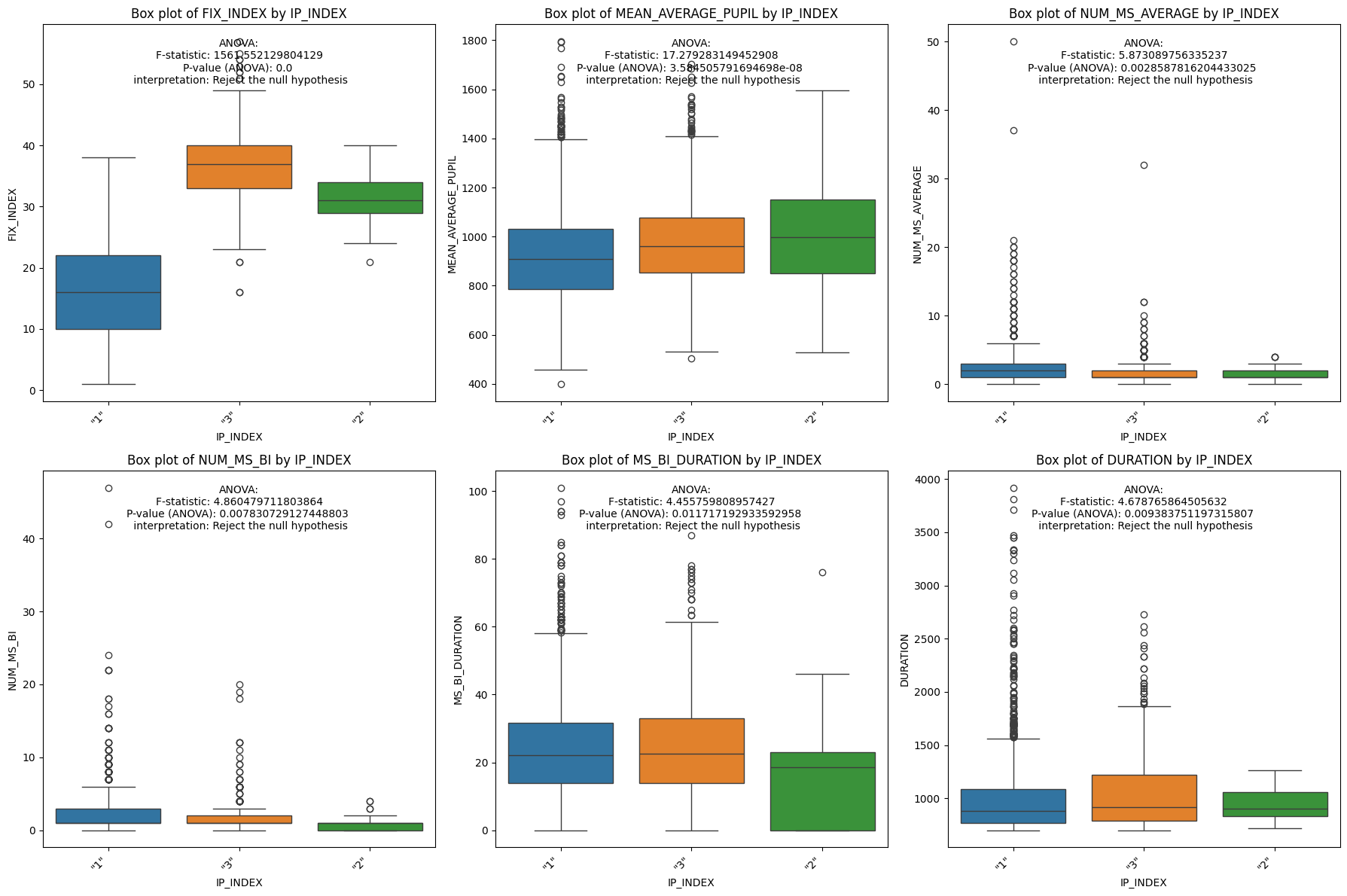}}
\caption{Comparison of numerical events across different interest periods (steps). Microsaccades and pupil size show significant differences, indicating the varying cognitive load and task complexity in each step.}
\label{fig:IPs}
\end{center}
\end{figure}

\begin{figure}[ht]

\begin{center}
\centerline{\includegraphics[width=0.9\columnwidth]{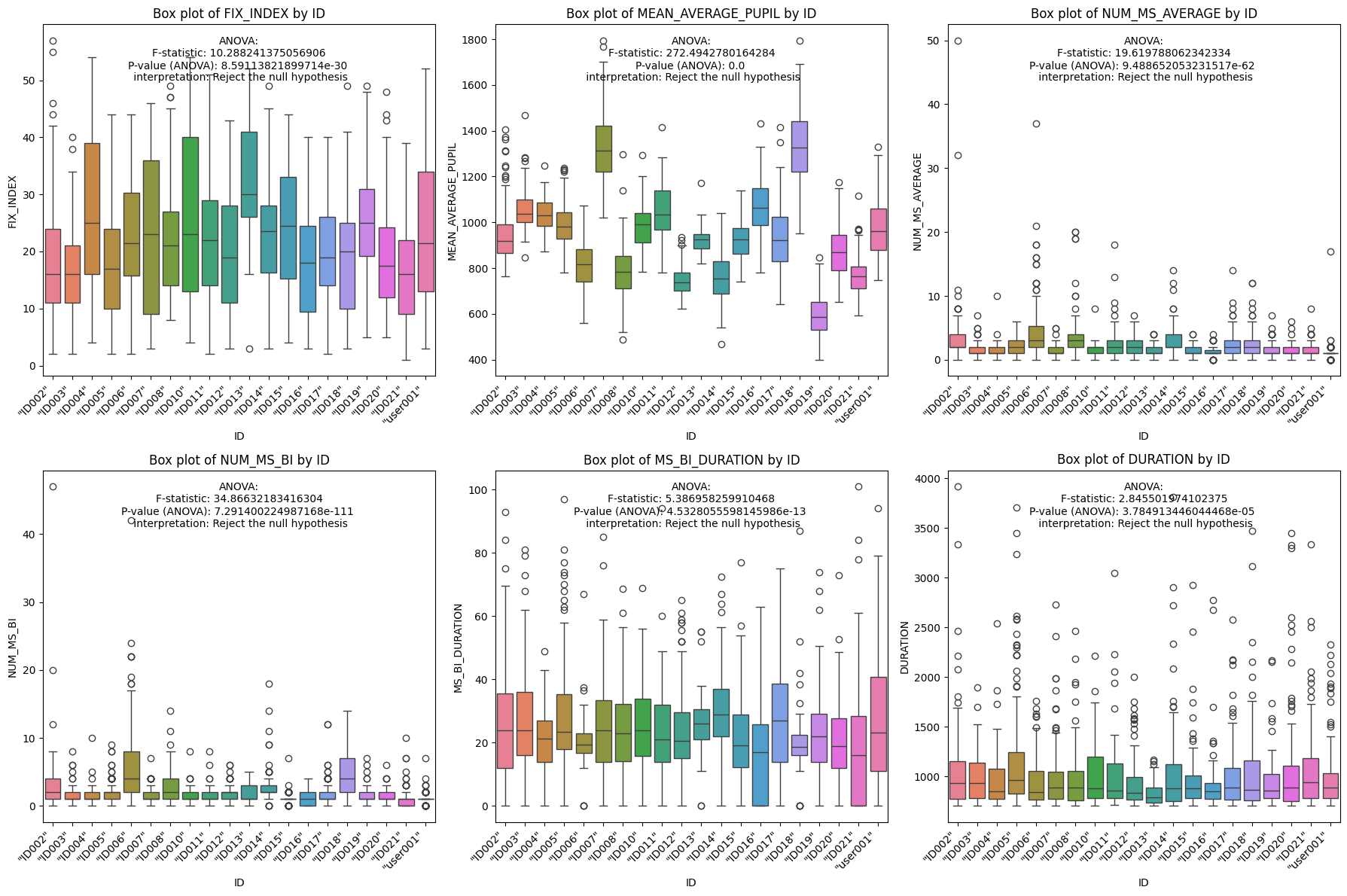}}
\caption{Variability in pupillary, microsaccadic, and fixational activities among participants. This underscores the individual differences in emotion perception and the challenge of creating a universal model.}
\label{fig:participants}
\end{center}
\end{figure}

\begin{table}[htbp]
\caption{ANOVA test results between numerical and categorical variables. The results highlight significant associations between various eye-tracking metrics and task-related factors.}
\centering
\resizebox{\textwidth}{!}{%
\begin{tabular}{|l|l|c|c|}
\hline
\textbf{Variable} & \textbf{Category} & \textbf{F-statistic} & \textbf{P-value (ANOVA)} \\
\hline
\multicolumn{4}{|c|}{\textbf{Emotions}} \\
\hline
Fixation index in trial & Fixation & 1.6851 & 0.1348 \\
Average pupil size & Pupil & 1.1715 & 0.3207 \\
Average of both eyes microsaccade's rate & Microsaccade & 1.2997 & 0.2611 \\
Binocular microsaccade's rate & Microsaccade & 2.0415 & 0.0700 \\
Binocular microsaccade average duration & Microsaccade & 0.9459 & 0.4500 \\
Fixation duration & Fixation & 1.3105 & 0.2566 \\
\hline
\multicolumn{4}{|c|}{\textbf{RoI Label}} \\
\hline
Fixation index in trial & Fixation & 2.7235 & 0.0123 \\
Average pupil size & Pupil & 10.4780 & \textbf{$1.7351 \times 10^{-11}$*} \\
Average of both eyes microsaccade's rate & Microsaccade & 2.7766 & 0.0108 \\
Binocular microsaccade's rate & Microsaccade & 4.2456 & \textbf{0.0003*} \\
Binocular microsaccade average duration & Microsaccade & 0.4616 & 0.8371 \\
Fixation duration & Fixation & 6.0402 & \textbf{$2.7859 \times 10^{-6}$*} \\
\hline
\multicolumn{4}{|c|}{\textbf{Face Region}} \\
\hline
Fixation index in trial & Fixation & 2.5302 & 0.0556 \\
Average pupil size & Pupil & 0.1862 & 0.9058 \\
Average of both eyes microsaccade's rate & Microsaccade & 2.4109 & 0.0651 \\
Binocular microsaccade's rate & Microsaccade & 7.6508 & \textbf{$4.3650 \times 10^{-5}$*} \\
Binocular microsaccade average duration & Microsaccade & 0.5778 & 0.6296 \\
Fixation duration & Fixation & 4.3748 & \textbf{0.0045*} \\
\hline
\multicolumn{4}{|c|}{\textbf{Participant ID}} \\
\hline
Fixation index in trial & Fixation & 2.9196 & 0.0542 \\
Average pupil size & Pupil & 0.1942 & 0.8235 \\
Average of both eyes microsaccade's rate & Microsaccade & 3.4939 & 0.0306 \\
Binocular microsaccade's rate & Microsaccade & 10.2293 & \textbf{$3.7856 \times 10^{-5}$*} \\
Binocular microsaccade average duration & Microsaccade & 0.0011 & 0.9989 \\
Fixation duration & Fixation & 3.7091 & 0.0247 \\
\hline
\multicolumn{4}{|c|}{\textbf{Interest Period Index}} \\
\hline
Fixation index in trial & Fixation & 10.8056 & \textbf{$1.1613 \times 10^{-31}$*} \\
Average pupil size & Pupil & 286.4408 & \textbf{0.0*} \\
Average of both eyes microsaccade's rate & Microsaccade & 20.3782 & \textbf{$1.4308 \times 10^{-64}$*} \\
Binocular microsaccade's rate & Microsaccade & 36.3224 & \textbf{$4.0576 \times 10^{-116}$*} \\
Binocular microsaccade average duration & Microsaccade & 5.4073 & \textbf{$3.7311 \times 10^{-13}$*} \\
Fixation duration & Fixation & 2.8619 & \textbf{$3.3758 \times 10^{-5}$*} \\
\hline
\end{tabular}
}
\label{tab:anova_results}
\end{table}

\begin{table}[htbp]
\centering
\caption{Chi-Square Test Results between categorical variables. The results indicate significant associations between categorical variables such as emotions, RoI labels, and face regions.}
\begin{tabular}{|l|l|c|c|}
\hline
\textbf{Variable} & \textbf{Category} & \textbf{Chi-Square} & \textbf{P-value} \\
\hline
\multicolumn{4}{|c|}{\textbf{Emotions}} \\
\hline
Emotions & RoI Label & 398.9696 & $4.5144 \times 10^{-66}$ \\
Emotions & Target Emotion & 11232.5746 & \textbf{0.0*} \\
Emotions & Face region & 286.1583 & $1.3057 \times 10^{-55}$ \\
\hline
\multicolumn{4}{|c|}{\textbf{RoI Label}} \\
\hline
RoI Label & Target Emotion & 68.7352 & \textbf{0.0001*} \\
RoI Label & Face region & 83284.0 & \textbf{0.0*} \\
\hline
\multicolumn{4}{|c|}{\textbf{Target Emotion}} \\
\hline
Target Emotion & Face region & 7.2468 & 0.7020 \\
\hline
\end{tabular}

\label{tab:chi_square_results}
\end{table}

In Step 2, participants intuitively sought to match the emotion word to the corresponding face. The emotion word and the position of the target face attracted the most attention, indicative of a memory effect. Specifically, the position of the target face received more focus, particularly for emotions like sadness, fear, and surprise, as depicted in Figure \ref{fig:FER_results}.

In Step 3, we noticed a new FER process where target faces ($M=40.9, SD=20.2$) received significantly more attention than non-target faces ($M=15.7, SD=11.8$) ($t(4318)=64.2, p<0.0001$). Non-target faces showing fear and surprise still attracted more attention than other non-target faces. Our results align well with previous studies that found participants tend to fixate longer on emotional faces, especially fearful and surprised ones, during daily communication.

Figure \ref{fig:eye_mouth_nose} depicts the relative distribution of eye, nose, and mouth regions of target and non-target faces in step 3 and all faces in step 1. The results reveal distinct attentional strategies for different emotions depending on the task requirements, whether free or emotion-guided observation. For instance, consistent with the findings of Polet et al. \cite{polet2022eye}, the eye region of surprised, fearful, and sad faces appears more crucial than that of other emotions in Step 1. Interestingly, this effect is even more pronounced when recognizing emotions in 3. Additionally, the mouth region is more critical for recognizing anger and disgust in 1 and, to a lesser extent, 3, compared to other emotions.

This comprehensive analysis highlights the intricate dynamics of eye movements in relation to emotion perception, underscoring the significance of detailed attention to microsaccades and fixation patterns across different stages of the emotion recognition task.

\begin{table}
\tbl{The performance metrics of models on Task 3.}
{\begin{tabular}{lcc} \toprule
Model & Mean Squared Error  & Spearman correlation \\ \midrule
XGBoost & 0.0628  & 0.2496* \\
LSTM & 0.0664 & 0.1980* \\ \bottomrule
\end{tabular}}
\tabnote{* $p < 0.05.$}
\label{tab:performance}
\end{table}

\begin{figure}
\centering
\subfloat[Participants]{%
\resizebox*{5cm}{!}{\includegraphics{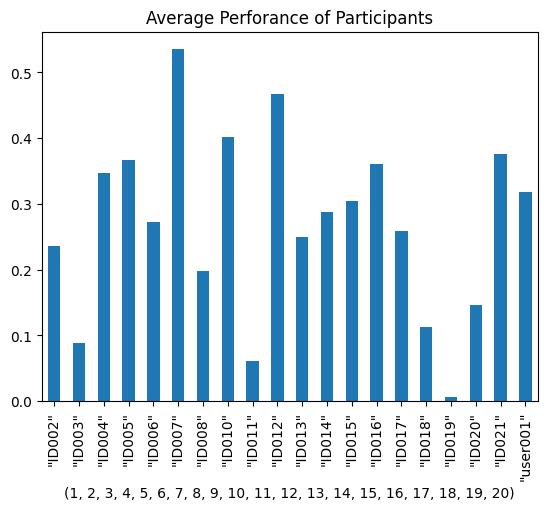}}}
\subfloat[Emotions]{%
\resizebox*{5cm}{!}{\includegraphics{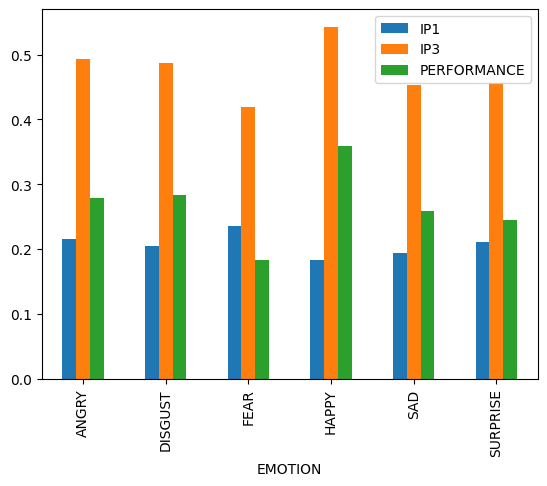}}}
\subfloat[Trials]{%
\resizebox*{5cm}{!}{\includegraphics{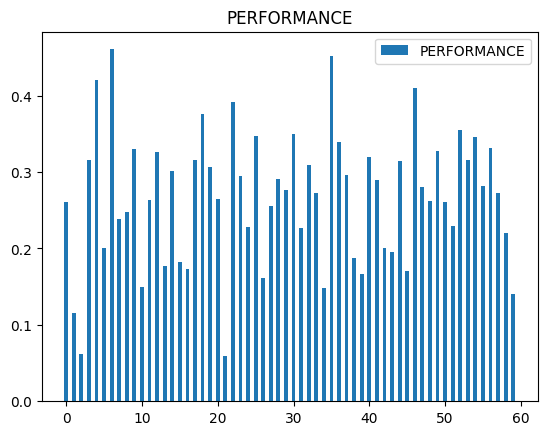}}}\hspace{5pt}
\caption{Performance analysis across participants, trials, and emotions. The analysis highlights individual differences in performance, emotional perception, and trial difficulty.} 
\label{fig:performance}
\end{figure}

%\vspace{-1em}
\section{Modeling}
We analyzed data from 20 participants over 54 trials, deriving 54 input-output pairs by averaging fixation events. We employed leave-one-out cross-validation due to the limited dataset size and focused on mean squared error (MSE) as our main performance metric.

\subsection{Hyperparameter Selection and Justification}
Our choice of hyperparameters for deep-learning models was informed by literature, grid search, and practical constraints, aiming to balance complexity and overfitting risk given our small dataset.

\subsection{Task 1}
Task 1 predicted dwell times for each face in Step 3 from Step 1's fixation data. This could improve instruction-free FER tasks reflecting everyday interactions. We predicted using spatial and temporal features of fixation, such as dwell time percentage and fixation counts, and emotion-related features, across four faces into a 3-layer Multilayer Perceptron (MLP) with 32-16-4 nodes. The model, trained for 500 epochs at a learning rate of 0.001, emphasized target face predictions by weighting them higher in the loss function.

\subsection{Task 2}
This task forecasted fixation dwell times using only visual features of faces. Employing a pre-trained VGG-Face model, we extracted 2622-dimensional embeddings as inputs for a 3-layer MLP (100-16-4 nodes). The larger first layer managed the high-dimensional data, reducing it progressively. The model ran for 1000 epochs with a learning rate of 0.001, focusing separately on data from Steps 1 and 3, prioritizing target face features.

\subsection{Task 3}
Task 3 aimed to predict individual user performance from eye movement events in Step 1. We calculated microsaccade rates and average pupil sizes from seven facial regions of interest for each emotion, using a 64-unit bidirectional LSTM for 100 epochs to capture temporal patterns. Performance prediction was handled by an XGBoost model, chosen for its effectiveness with structured data.

\subsection{Baseline Model}
The baseline model assumed the target face attracted the most attention, allocating the longest viewing time to it and dividing the remainder equally among other faces. This strategy provided a dwell time distribution of 0.50 for the target and 0.1666 for non-target faces in Tasks 1 and 2's Step 3, and equal times in Task 2's Step 1.

\section{Results}
Results indicated successful dwell time predictions with low MSE rates in both tasks. Task 1 showed that temporal features outperformed spatial ones, suggesting the importance of temporal data in modeling complex FER tasks. Task 2 confirmed the difficulty of emotion-related task predictions, with Step 3 posing greater challenges than Step 1.

\begin{table}[]
\centering
\caption{Average MSE results for predicting the dwell time of Tasks 1 and 2. The results indicate that temporal and spatiotemporal features significantly enhance prediction accuracy compared to baseline and spatial features alone.}
\begin{center}
\begin{small}
\begin{sc}
    \begin{tabular}{l ll}
        \toprule
        \multicolumn{3}{c}{\textsc{Task 1}}\\
        
        \textsc{features} & \textsc{all} & {\textsc{target}} \\ 
        \midrule
        Baseline & 0.0164 & 0.0060 \\
        Spatial & 0.0134 & 0.0066 \\
        Temporal & 0.0053 & 0.0030 \\ 
        Spatiotemporal & \textbf{0.0046} & \textbf{0.0024} \\ 
        \midrule
        \multicolumn{3}{c}{\textsc{Task 2}}\\
        
        \textsc{features} & \textsc{Step 1} & \textsc{Step 3}\\
        \midrule
        Baseline & 0.0152 & 0.0164 \\
        Face embeddings & \textbf{0.0065} & \textbf{0.0077}\\
        \bottomrule
    \end{tabular}%
    \end{sc}
    \end{small}
    \end{center}
    \vskip -0.1in
    \vspace{-1em}
\label{tab:model_results}
\end{table}

Table \ref{tab:performance} displays the results for performance detection models as discussed in the previous section. We utilized mean squared error as a metric for evaluating the results, although it may not provide a comprehensive understanding of the model's performance quality. Therefore, we employed correlation as an evaluation metric to demonstrate the degree of correlation between predicted and true performances. Contrary to our expectations, the BiLSTM sequential model performed worse than the XGBoost model. However, neither of the models exhibited strong performance. As demonstrated in the statistical analysis, developing a single model capable of accurately predicting performance for all users is a highly challenging task.

%\vspace{-0.5em}
\section{Discussion and Conclusion}
\label{sec:conclusion}

We adapted Russell et al.'s (2021) instructionless Facial Expression Recognition (FER) task to delve deeper into the FER process. These modifications facilitated extensive statistical analysis and revealed crucial disparities in processing various emotions.

By dividing the emotion processing into two steps—free-viewing and emotion grounding we gained deeper insights into emotion perception. We demonstrated the ability to predict individuals' performance solely from features observed during the free-viewing steps, aligning with eye-gaze strategies for facial emotion perception.

Our findings indicate that gaze events, particularly temporal features, can predict FER performance by merely observing faces. We also forecasted the fixation duration of FER tasks based on facial visual features, aiding in the assessment of trial difficulty. Notably, we predicted emotion perception accuracy from free face viewing, marking progress towards lightweight emotion recognition assessments not dependent on language skills.

As depicted in Figure \ref{fig:performance}, the difficulty levels of trials vary significantly. The work conducted in task 2 is instrumental in evaluating the difficulty level of any given trial. Additionally, our statistical analysis revealed substantial variations in participants' performances, reflected in their fixational, microsaccadic, and pupillary activities. Understanding these differences requires approximating the average participant, which the work in task 1 provides. Therefore, tasks 1 and 2 furnish information about trial difficulty and average fixation distribution, thereby informing performance. Task 3 predicts individual performance based on free face viewing, amalgamating all this information to offer a comprehensive emotion analysis of a user and enabling the modeling of users solely from their data during free face processing. This provides a naturalistic platform that can implicitly learn from users while they engage in their natural tasks.

Moreover, we introduced a standardized tool for FER datasets, enhancing the comparability of results. Overall, our work offers insights for FER research and could shape the development of more naturalistic emotion recognition assessments.

While the temporal modeling of eye gaze strategies presents a promising avenue, further investigation is warranted. Specifically, the extraction of compatible events and an increase in data collection could significantly enhance performance. Another factor to consider is the challenge posed by having four faces on the screen simultaneously, which can make it difficult to distinguish between regions of interest across different faces. Designing a new setting where each face is presented separately could address this issue and provide more valid temporal data on gaze strategies. Additionally, incorporating participant annotations of perceived emotions could enrich the analysis by providing a more valid ground truth for modeling emotion perception.

In conclusion, our work advances FER by exploring new paradigms and models. Predicting FER performance from free-viewing eye movements offers a pathway for efficient and ecologically valid emotion perception assessments. We anticipate that our work will inspire further research and foster improved tools and methodologies for studying human emotion perception.

%\bibliographystyle{apacite}

% Your content here

\printbibliography

\end{document}